\documentclass[a4paper,11pt]{article}

\usepackage{jheppub} 

\newcommand{\beq}{\begin{equation}}
\newcommand{\beqa}{\begin{eqnarray}}
\newcommand{\eeq}{\end{equation}}
\newcommand{\eeqa}{\end{eqnarray}}

\newcommand{\simgt}{\lower.5ex\hbox{$\; \buildrel > \over \sim \;$}}
\newcommand{\simlt}{\lower.5ex\hbox{$\; \buildrel < \over \sim \;$}}

\newcommand{\bd}[1]{\mbox{\boldmath $#1$}}

\title{\boldmath Weighing the Light Gravitino Mass with Weak Lensing Surveys}

\author[a]{Ayuki Kamada,}
\author[b]{Masato Shirasaki}
\author[a,b]{and Naoki Yoshida}
\affiliation[b]{Department of Physics, University of Tokyo, Tokyo 113-0033, Japan}
\affiliation[a]{Kavli IPMU (WPI), University of Tokyo, Chiba 277-8583, Japan}

\emailAdd{ayuki.kamada@ipmu.jp}
\emailAdd{masato.shirasaki@utap.phys.s.u-tokyo.ac.jp}

\abstract{
We explore the discovery potential of light gravitino mass $m_{3/2}$ 
by combining future cosmology surveys and collider experiments. 
The former probe the imprint
of light gravitinos in the cosmic matter density field, 
whereas the latter search signatures of a supersymmetry breaking 
mechanism.
Free-streaming of light gravitinos suppresses the density fluctuations 
at galactic and sub-galactic length scales, where weak gravitational 
lensing can be used as a powerful probe. 
We perform numerical simulations of structure 
formation to quantify the effect.
We then run realistic ray-tracing simulations of gravitational lensing 
to measure the cosmic shear in models with light gravitino. 
We forecast the possible reach of future wide-field surveys 
by Fisher analysis; the light gravitino mass can be determined with 
an accuracy of $m_{3/2}=4\pm1 $\,eV by a combination of the Hyper Suprime 
Cam survey and cosmic microwave background 
anisotropy data obtained by Planck satellite. 
The corresponding accuracy to be obtained by the future 
Large Synoptic Survey Telescope is $\delta m_{3/2}=0.6 $\,eV.
Data from experiments at Large Hadron Collider at $14$\,TeV
will provide constraint at $m_{3/2} \simeq 5$\,eV in the minimal 
framework of gauge-mediated 
supersymmetry breaking (GMSB) model. 
We conclude that a large class of the GMSB model can 
be tested by combining the cosmological observations 
and the collider experiments.
}

\begin{document} 
\maketitle
\flushbottom

\section{Introduction}
Supersymmetry (SUSY) is one of the most attractive candidates of 
physics beyond the standard model.
Minimal supersymmetric extension of the standard model 
(MSSM, see e.g.\,\cite{Martin:1997ns}) with 
$\sim {\mathcal O}({\rm TeV})$ SUSY particles
can possibly address several important issues in the standard model, 
such as the large hierarchy between the electro-weak scale and
the grand unification scale, 
the existence of dark matter, and the origin of the cosmic baryon number.
The MSSM is also known to achieve successful grand unification of the 
standard model gauge couplings at some high energy.

The null-detection of the SUSY particles so far suggests that 
SUSY is broken at some energy scale and mediated to MSSM via some messenger.
Several mechanisms are proposed as the messenger such as gravity-mediated, 
anomaly-mediated, and gauge-mediated SUSY breaking (GMSB) models. 
GMSB models generally evade the flavor changing neutral current 
problem and the CP-problem,
and thus they are thought to be the most interesting models.

Supergravity (SUGRA) as an extension of the global SUSY to the local one 
involves the superpartner 
of the graviton, which is referred to gravitino.
The gravitino has helicity $3/2$ and obtains the mass via the 
super-Higgs mechanism:
\beqa
\label{eq:gravitinomass}
m_{3/2} = \frac{| \langle F \rangle |}{\sqrt{3} M_{\rm pl}} \,,
\eeqa
with the vacuum expectation value of $F$-term $\langle F \rangle$
and the reduced Planck mass $M_{\rm pl} \simeq 2.43 \times 10^{18}$\,GeV.
The gravitino is produced in the thermal bath immediately after 
the reheating of the Universe
through the $F$-term suppressed interaction of goldstino component 
with spin $\pm 1/2$.


In the GMSB models, the gravitino mass is predicted to be in the range of 
$m_{3/2} \simeq {\rm eV}-{\rm keV}$.
The small $F$-term allows the gravitino to be in the thermal equilibrium 
until the decoupling of others SUSY particles.
When the gravitino is decoupled from the thermal bath, it begins to stream 
freely and contributes as a ``diffuse'' matter component of the Universe.  
The gravitino free-streaming imprints characteristic features on the
matter power spectrum, which are expected to be probed by observations
of large-scale structure.
For example, the current constraint of $m_{3/2} < 16$\,eV is obtained by 
measuring the Ly-$\alpha$ flux power spectra that essentially probe 
the distribution of the inter-galactic medium at high redshifts
\,\cite{Viel:2005qj}. We note that the constraint is based on the
crucial assumption that the distribution of the inter-galactic
neutral gas traces the distribution of underlying dark matter
even at nonlinear length scales.
Gravitational lensing provides a direct physical mean of probing
the distribution of total matter.
For example, it has been suggested that cosmic microwave background 
lensing has a potential to probe the gravitino mass of $m_{3/2} \simeq 1$\,eV 
in future experiments \,\cite{Ichikawa:2009ir}.

While the cosmological observations place an upper bound on the 
gravitino mass, the terrestrial collider experiments such as 
on-going Large Hadron Collider (LHC) 
give a lower bound through signatures of other SUSY particles 
(see Section\,\ref{sec:massspectrum}).
In the present paper, we show that essentially all the interesting range of 
the gravitino mass can be probed by combining the up-coming LHC 
run at $14$\,TeV and 
the near future weak lensing surveys by the Subaru Hyper Suprime-Cam (HSC) and 
the Large Synoptic Survey Telescope (LSST).

Gravitational lensing is one of the powerful tools to probe directly 
the matter distribution in the Universe.
The coherent pattern of image distortion by weak lensing is called 
cosmic shear. Cosmic shear in principle can be induced by any foreground 
mass distribution along 
the line of sight regardless of its dynamical state or luminosity.
Cosmic shear signals have been detected with high significance levels,
and constraints on some basic cosmological parameters have been derived 
\cite{Bacon:2002va,Hamana:2002yd, Benjamin:2007ys, Kilbinger:2012qz}.
Upcoming weak lensing surveys such as HSC will cover a wide area 
extending more than 
a thousand square degrees. The surveys will also probe the matter distribution
at $\sim$ Mpc scale most accurately, where
the imprints of the gravitino can be detected. 
It is therefore important and timely to study the effect of the 
light gravitino on cosmic shear.
To this end, we run a set of cosmological $N$-body simulations to 
follow the nonlinear evolution of the matter density fluctuations 
with the imprints of the gravitino free-streaming.
We then perform accurate ray-tracing simulations of gravitational lensing.
We show that the cosmic shear is indeed a promising probe of the existence
and the mass of the light gravitino. 

The rest of the paper is organized as follows.
In Section\,\ref{sec:massspectrum}, we introduce the basics of the GMSB model.
In particular, we clarify the relation between the gravitino mass and the masses of other SUSY particles. 
In Section\,\ref{sec:linearpower}, we discuss the linear evolution of the primordial density perturbation
under the effect of the light gravitino.
We present the resulting linear matter power spectra, 
which provides the initial conditions for our cosmological $N$-body simulations.
In Section\,\ref{sec:cosmicshear}, we describe our simulation set-ups.
In Section\,\ref{sec:results}, we measure the cosmic shear power spectra from the simulations and 
forecast the discovery potential of the light gravitino in the future weak lensing surveys.
The final section is devoted to the concluding remarks.

\section{SUSY particle masses in the GMSB model}
\label{sec:massspectrum}
In the GMSB models\,\cite{Dine:1981gu, Nappi:1982hm, AlvarezGaume:1981wy, Dine:1993yw, Dine:1994vc, Dine:1995ag}, 
the SUSY breaking is mediated from the hidden sector to the MSSM sector 
via some messenger fields that are charged under the standard model gauge group.
The gaugino and the sfermion masses are induced 
by the one-loop and the two-loop diagrams, respectively, at the leading order.
Note that the gaugino and sfermion mass spectrum generically depends on the charge assignment to the messenger fields and that
inadequate charge assignment might ruin the success of MSSM 
in the grand unification of the gauge couplings.
A popular choice is to set messenger fields in complete multiplets 
of the $SU(5)$ global/gauge symmetry.
In the rest of this section, we consider specifically one of such models,
the so-called minimal GMSB model.

The minimal GMSB model has the superpotential of
\beqa
W = (\lambda S + M) \sum_{n=1}^{N_{5}} \Phi_{n} {\bar \Phi}_{n} \,,
\eeqa 
where $S$ is the goldstino superfields and $M$ is the messenger mass.
The $F$-term of the goldstino superfields develops the vacuum expectation 
values $\langle F \rangle$,
and the $N_{5}$ pairs of messenger superfields $\Phi_{n}$ 
and ${\bar \Phi}_{n}$ ($n=1, ..., N_{5}$) form the multiplets of ${\bf 5}$ and ${\bar {\bf 5}}$ of $SU(5)$.
In the minimal GMSB model, the gaugino mass is given by,
\beqa
\label{eq:gauginomass}
M_{a} = \frac{g_{a}^{2}}{16\pi^{2}} \Lambda N_{5} g(x),
\eeqa
where the index $a\,(=1,2,3)$ corresponds to the standard model gauge group 
$U(1)_{Y} \times SU(2)_{L} \times SU(3)_{C}$,
and $g_{a}$ denotes the standard model gauge coupling.
We normalize $g_{1}$ and $g_{2}$ such that $g_{1} = \sqrt{5/3} \, g'$ and $g_{2} = g$ 
with the conventional electro-weak
gauge couplings $g$ and $g'$ ($e = g \sin \theta_{W} = g' \sin \theta_{W}$, $e$: positron charge, $\theta_{W}$: Weinberg angle).
The messenger scale $\Lambda$ is defined by,
\beqa
\Lambda =  \left| \frac{\lambda \langle F \rangle }{M} \right| \,.
\eeqa
The function $g(x)$ is given by,
\beqa
g(x)=\frac{1}{x^{2}} (1+x) \ln (1+x) + (x \rightarrow -x) \,,
\eeqa
and its argument is the dimensionless parameter $x = \left| \Lambda/M \right|$.
The sfermion mass squared is given by
\beqa
\label{eq:sfermionmass}
m^{2}_{\phi_{i}} = 2 \Lambda^{2} N_{5} \sum_{a=1}^{3} C_{a}(i) \left( \frac{g_{a}^{2}}{16\pi^{2}} \right)^{2} f(x)\,,
\eeqa
where the index $i$ denotes fermion flavour and $C_{a}(i)$ 
is the Casimir invariant.
The function $f(x)$ is given by,
\beqa
f(x)=\frac{1+x}{x^{2}} \left[ \ln (1+x) -2 {\rm Li}_{2} \left( \frac{x}{1+x} \right) + \frac{1}{2} {\rm Li}_{2} \left( \frac{2x}{1+x} \right) \right]  + (x \rightarrow -x) \,,
\eeqa
with the dilogarithm function ${\rm Li}_{2}(x)$.
In practice, we use the public code {\tt softsusy}\,\cite{Allanach:2001kg} 
to calculate the mass spectrum of SUSY particles numerically. 
The calculations take into account the renormalization group running 
of SUSY particle masses.
The gravitino mass (eq.\,(\ref{eq:gravitinomass})) can be written in terms of the GMSB variables,
\beqa
\label{eq:gravitinomassp}
m_{3/2} = \frac{\Lambda M}{\sqrt{3} M_{\rm pl} |\lambda|} = \frac{\Lambda^{2}}{\sqrt{3} M_{\rm pl} |\lambda| x} \,.
\eeqa

From eqs.\,(\ref{eq:gauginomass}), (\ref{eq:sfermionmass}) and (\ref{eq:gravitinomassp}), we can see that the SUSY particle masses are proportional to the square root of the gravitino mass,
$M_{a},\,m_{\phi_{i}} \propto \Lambda \propto \sqrt{m_{3/2}}$. Therefore, collider experiments
with higher energies can be used generally to search for signatures 
of heavier SUSY particles, 
which in turn give information on gravitinos with relatively larger masses.
Note that, in high-energy collision of the standard model particles, the direct product 
is not gravitino with gravitational interaction, 
but other SUSY particles with gauge interaction.
For example, in proton-proton collision experiments 
at the LHC, the colored SUSY particles (i.e. gluino and squarks) 
are important and directly related to the discovery potential for SUSY particles.
Lighter gravitinos are associated with lighter colored SUSY particles
that can be searched even with the current generation experiments.

In order to derive conservative constraints on the gravitino mass 
from the collider experiments, 
we consider models with maximal $N_{5}$ and $|\lambda| x$.
For the successful grand unification of the gauge couplings, the number of messenger $N_{5}$ 
needs to be at most five, $N_{5} \le 5$.
We set $N_{5}=5$ as a conservative choice. 
The LHC current and future reach for the GMSB models is studied in detail 
in\,\cite{Baer:1998ve, Baer:2000pe, Aad:2009wy, Ludwig:2010kt, Nakamura:2010faa, 
ATLAS:2012ag, Aad:2012rt, ATLAS:2012ht, ATLAS-CONF-2013-026}. 
Let us consider two focus points specifically, as summarized 
in Table\,\ref{table:lowerbound}.
The ATLAS collaboration sets the lower bound on $\Lambda>51$\,TeV with $M=250$\,TeV 
and $N_{5}=3$ (${\bf 10}+{\bar {\bf 10}}$ of $SU(5)$) fixed 
from the events with at least one tau lepton and no light lepton 
in $21\,{\rm fb}^{-1}$ of LHC $8$\,TeV run\,\cite{ATLAS-CONF-2013-026}.
This can be interpreted as a lower bound on the gravitino mass $m_{3/2}>3$\,eV 
through eq.\,(\ref{eq:gravitinomassp}), 
for the assumed perturbative coupling $|\lambda|<1$.

It is expected that the $\Lambda=80$\,TeV is accessible even 
for $N_{5}=5$ with the use of the multi-lepton modes in 
about $15\,{\rm fb}^{-1}$ of LHC $14$\,TeV run\,\cite{Nakamura:2010faa}. 
The stability of the SUSY breaking vacuum requires $|\lambda| x>1.4$\,\cite{Hisano:2007gb}.
\footnote{Considering the thermal transition of the SUSY breaking vacuum leads to more stringent constraint on $|\lambda| x$\,\cite{Hisano:2008sy}.
Here, we consider only quantum (zero-temperature) transition to be conservative.}
An exciting implication of this is that virtually 
all of the GMSB models with $m_{3/2}<5$\,eV can be probed 
in $15\,{\rm fb}^{-1}$ of LHC $14$\,TeV run.
Later in section 4, we show that the future weak lensing survey 
can determine the gravitino mass 
with an accuracy of $m_{3/2}=4\pm1$\,eV, combined with the CMB 
anisotropy measurement by Planck satellite.
We suggest that there is a good chance to test fundamentally the GMSB 
models in the near future.


\begin{table}[tb]
  \begin{center}
  \begin{tabular}{|c||c|c|c|c|} \hline
  focus point & fixed GMSB parameters & LHC & $\Lambda$ & $m_{3/2}$ \\ \hline \hline
  current  & $M=250$\,TeV,\,$N_{5}=3$,\,$|\lambda|=1$ & $21\,{\rm fb}^{-1}$\,at\,$\sqrt{s}=8$\,TeV & $\Lambda=51$\,TeV & $m_{3/2}=3$\,eV \\ \hline
  future & $|\lambda| x=1.4$,\,$N_{5}=5$ & $15\,{\rm fb}^{-1}$\,at\,$\sqrt{s}=14$\,TeV & $\Lambda=80$\,TeV & $m_{3/2}=5$\,eV \\ \hline    
  \end{tabular}
  \end{center}
  \caption{
  Summary of the focus points for the GMSB model described in the text. 
  The current focus point corresponds to the current lower bound on $\Lambda$ reported in\,\cite{ATLAS-CONF-2013-026}.
  In the future focus point, the GMSB parameters are set 
  such that they minimize the gravitino mass for fixed $\Lambda$ while stabilizing the SUSY breaking vacuum.
  The future LHC reach is taken from\,\cite{Nakamura:2010faa}.
   }
   \label{table:lowerbound}
\end{table}

\section{Linear evolution of density perturbations with light gravitino}
\label{sec:linearpower}
The light gravitino is in thermal equilibrium immediately 
after the reheating of the Universe 
unless the reheating temperature is extremely low\,\cite{Moroi:1993mb}.
When the cosmic temperature drops below the other SUSY particle masses, 
the decay and inverse-decay processes
that have been keeping the thermal equilibrium between the light gravitino 
and the thermal bath, become inefficient.
Then the light gravitino particles 
begins to stream freely with the momenta following 
the Fermi-Dirac distribution.

The gravitino contribution to the cosmic energy density is given by,
\beqa
\Omega_{3/2}h^{2} = 0.13 \left( \frac{g_{3/2}}{2} \right) \left( \frac{m_{3/2}}{100\,{\rm eV}} \right) \left( \frac{g_{*s 3/2}}{90} \right)^{-1}
\eeqa
where $g_{*s 3/2}$ is the effective massless degrees of freedom for the cosmic entropy at the time of gravitino decoupling.
The exact value of $g_{*s 3/2}$ depends on the mass spectrum of the SUSY particles (e.g. $\Lambda$)\,\cite{Pierpaoli:1997im, Ichikawa:2009ir}.
However, its weak dependence allows us to fix $g_{*s 3/2} = 90$ without changing our results by more than $5$\,\%.
It should be noted that the effective internal degrees of freedom of gravitino is not $g_{3/2}=4$, but $g_{3/2}=2$.
This is because only goldstino component (spin $\pm1/2$) can interact with the thermal bath through the $1/ \langle F \rangle$ suppressed interactions.

From the above formula, we find that the light gravitino with 
$m_{3/2} \lesssim 100$\,eV (of our interest here) cannot account 
for the cosmological dark matter mass density.
We assume that some cold and stable particle makes up the rest of dark matter, i.e.,
\beqa
\Omega_{\rm dm} = \Omega_{\rm cs} + \Omega_{3/2} \,.
\eeqa
Such a cold and stable particle can be the QCD axion\,\cite{Preskill:1982cy, Abbott:1982af, Dine:1982ah} or the composite baryons in the SUSY breaking, 
or can also be generated in the messenger sector in models with strongly coupled 
low scale gauge mediation\,\cite{Dimopoulos:1996gy, Fan:2010is, Yanagida:2010zz, Yanagida:2012ef}.

\begin{figure}[t]
 \begin{center}
 \includegraphics[width=0.6\linewidth]{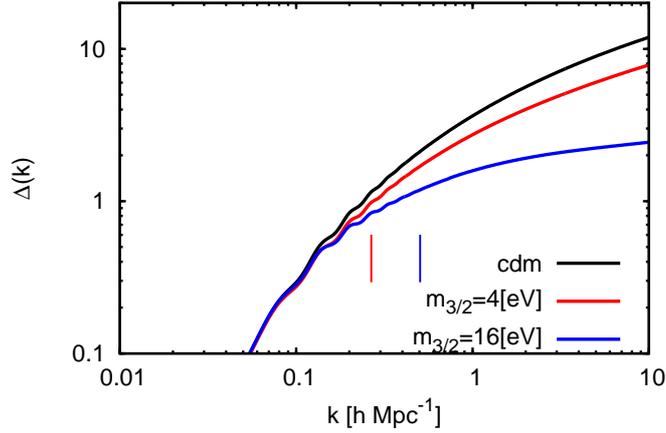}
 \caption{
 Linear dimensionless matter power spectra for $m_{3/2}=0\,(\rm cdm),\,4,\,{\rm and}\,16$\,eV.  
 We adopt basic cosmological parameters given by the Planck mission first year results \cite{Ade:2013zuv}.
 The vertical lines represents the cut-off scales of eq.\,(\ref{eq:Jeans}).
 }
 \label{fig:delta}
 \end{center}
\end{figure}

The free-streaming of light gravitino affects the evolution of primordial 
density perturbations in a similar manner as the standard model neutrinos do. 
We discuss the similarity and indeed the degeneracy of the
effects of the light gravitino and the standard model neutrinos later in Section\,\ref{sec:results}.
The suppression owing to free-streaming occurs below a cut-off scale 
that is characterized
by the Jeans scale at the matter-radiation equality $a_{\rm eq}$\,\cite{Kamada:2013sh}:
\beqa
\label{eq:Jeans}
k_{\rm J} = a \sqrt{\frac{4 \pi G \rho_{\rm M}}{\langle v^{2} \rangle}} {\bigg |}_{a=a_{\rm eq}} 
\simeq 0.86 \,{\rm Mpc}^{-1} \left( \frac{g_{3/2}}{2} \right)^{-1/2} \left( \frac{m_{3/2}}{100\,{\rm eV}} \right)^{1/2} \left( \frac{g_{*s 3/2}}{90} \right)^{5/6} \,,
\eeqa
where $G$ is the Newton's constant and $\rho_{\rm M}$ is the total matter density of the Universe.
The mean square velocity $\langle v^{2} \rangle$ is evaluated over the whole dark matter 
mass distribution function $f_{\rm dm}(v)$ ($\int d^{3}v f_{\rm dm}(v) = \rho_{\rm dm}$).
It means $\langle v^{2} \rangle = f_{3/2} \, \langle v^{2} \rangle_{3/2}$ effectively,
where $f_{3/2}$ is the gravitino density fraction ($f_{3/2} \equiv \Omega_{3/2} / \Omega_{c}$) 
and $\langle v^{2} \rangle_{3/2}$ is the mean square velocity of the gravitino particles. The resultant linear matter power spectrum is shown and compared with that 
of the standard $\Lambda$CDM model in Figure\,\ref{fig:delta}.

Before discussing the details of the matter power spectrum,
let us briefly consider the effect of some non-standard thermal 
history of the Universe.
The overall influence of the light gravitino on the cosmic expansion can be 
basically characterized by one parameter, the gravitino mass $m_{3/2}$.
This is because the gravitino temperature of the Fermi-Dirac distribution 
can be related to the observed CMB temperature through 
entropy conservation, i.e. $g_{*s 3/2} \simeq 90$.
However, non-standard thermal history with, e.g., entropy 
production\,\cite{Ibe:2010ym}, can change $g_{*s 3/2}$ 
drastically to $g_{*s 3/2}\simeq{\mathcal O}(1000)$.
Then cosmological constraint on $m_{3/2}$ can be significantly altered,
or the constraint needs to be re-interpreted within a suitable class of models.
In the following discussion, we consider the standard thermal 
history with $g_{*s 3/2} = 90$, and hence $m_{3/2}$ is the single model parameter.

We calculate the evolution of the linear density perturbations $\delta$
by modifying {\tt CAMB}\,\cite{Lewis:1999bs} suitably.
In Figure\,\ref{fig:delta}, we plot the dimensionless matter power spectra 
$\Delta (k)$ defined by
\beqa
\langle \delta({\bf x}) \delta({\bf y}) \rangle = \int d \ln k \, \Delta (k) \, e^{i {\bf k} \cdot ({\bf x} - {\bf y})}\,,
\eeqa
for mixed dark matter models with $m_{3/2}=0\,(\rm cdm),\,4,\,{\rm and}\,16$\,eV. 
We adopt the cosmological parameters of the Planck mission 
first year results \cite{Ade:2013zuv}.
The free-streaming effect appears clearly at small length 
scales (eq.\,\ref{eq:Jeans})
but the suppression below the cut-off scale is more significant for models 
with heavier gravitino 
(compare $m_{3/2}=4$\,eV and $16$\,eV in Figure\,\ref{fig:delta}).
This is because larger $m_{3/2}$ gives a larger fractional contribution 
to the total matter density as $f_{3/2} \propto m_{3/2}$.
We thus expect that models with heavy gravitino
can be constrained by observations of large-scale structure of the universe.

\section{Weak gravitational lensing}
\label{sec:cosmicshear}
\subsection{Lensing power spectrum}
We summarize basics of gravitational lensing by large-scale structure.
When one denotes the observed position of a source object as $\bd{\theta}$ 
and the true position as $\bd{\beta}$,
one can characterize the distortion of image of a source object by the 
following 2D matrix:
\beqa
A_{ij} = \frac{\partial \beta^{i}}{\partial \theta^{j}}
           \equiv \left(
\begin{array}{cc}
1-\kappa -\gamma_{1} & -\gamma_{2}  \\
-\gamma_{2} & 1-\kappa+\gamma_{1} \\
\end{array}
\right), \label{distortion_tensor}
\eeqa
where $\kappa$ is convergence and $\gamma$ is shear.
In weak lensing regime (i.e. $\kappa, \gamma \ll 1$), 
each component of $A_{ij}$ can be related to
the second derivative of the gravitational potential $\Phi$ \cite{Bartelmann:1999yn,Munshi:2006fn} as
\beqa
A_{ij} &=& \delta_{ij} - \Phi_{ij}, \label{eq:Aij} \\
\Phi_{ij}  &=&\frac{2}{c^2}\int _{0}^{\chi}{\rm d}\chi^{\prime} g(\chi,\chi^{\prime}) 
\frac{\partial^2}{\partial x_{i}\partial x_{j}}\Phi[r(\chi^{\prime})\bd{\theta},\chi^{\prime}], \label{eq:shear_ten}\\	
g(\chi,\chi^{\prime}) &=& \frac{r(\chi-\chi^{\prime})r(\chi^{\prime})}{r(\chi)}
\eeqa
where $\chi$ is comoving distance, $r(\chi)$ is angular diameter distance, 
and $x_{i}=r\theta_{i}$ represents physical distance.
By using the Poisson equation, one can relate the 
convergence field to the matter overdensity field $\delta$
\cite{Bartelmann:1999yn,Munshi:2006fn}.
Weak lensing convergence field is then given by
\beqa
\kappa(\bd{\theta},\chi)= \frac{3}{2}\left(\frac{H_{0}}{c}\right)^2 \Omega_{\rm m0}
\int _{0}^{\chi}{\rm d}\chi^{\prime} g(\chi,\chi^{\prime}) 
\frac{\delta[r(\chi^{\prime})\bd{\theta},\chi^{\prime}]}{a(\chi^{\prime})}. \label{eq:kappa_delta}
\eeqa

In this paper, we use the convergence power spectrum to
constrain the gravitino mass.
With the flat sky approximation, which is sufficient for 
angular scales of our interest, 
the Fourier transform of convergence field is defined by
\beqa
\kappa(\bd{\theta})= \int \frac{{\rm d}^2 \ell}{(2\pi)^2}e^{i\bd{\ell}\cdot\bd{\theta}}\tilde{\kappa}(\bd{\ell}).
\eeqa
The power spectrum of the convergence field $P_{\kappa}$ is defined by 
\beqa
\langle \tilde{\kappa}(\bd{\ell}_{1}) \tilde{\kappa}(\bd{\ell}_2)\rangle
=(2\pi)^2 \delta_{D}(\bd{\ell}_{1}-\bd{\ell}_{2}) P_{\kappa}(\ell_1),
\eeqa
where $\delta_{D}(\bd{\ell})$ is the Dirac delta function.
By using Limber approximation \cite{Limber:1954zz,Kaiser:1991qi}
and eq.~\eqref{eq:kappa_delta}, 
we obtain the convergence power spectrum as 
\beqa
P_{\kappa}(\ell) &=& \int_{0}^{\chi_s} {\rm d}\chi \frac{W(\chi)^2}{r(\chi)^2} 
P_{\delta}\left(k=\frac{\ell}{r(\chi)},z(\chi)\right)
\label{eq:kappa_power},
\eeqa
where $P_{\delta}(k)$ is the three dimensional matter power spectrum, $\chi_s$ is comoving distance of source galaxies and $W(\chi)$ is the lensing weight function defined as
\beqa
W(\chi) = \frac{3}{2}\left(\frac{H_{0}}{c}\right)^2 \Omega_{\rm m0}
\frac{r(\chi_s-\chi)r(\chi)}{r(\chi_s)}(1+z(\chi)).
\eeqa

The non-linear gravitational growth of $P_{\delta}(k)$ significantly affects 
the amplitude of convergence power spectrum for the angular scales less than 1 degree
\cite{Jain:1999ir, Hilbert:2008kb, Sato:2009ct}.
Typical weak lensing surveys are
aimed at measuring the cosmic shear at 
angular scales larger than a few arcmin, corresponding to a few mega-parsec.
Therefore, accurate theoretical prediction of non-linear matter power 
spectrum is essential
to derive cosmological constraints from weak lensing power spectrum.
Several analytic models are available that 
accurately predict the non-linear evolution of $P_{\delta}(k)$ for 
the standard $\Lambda$CDM universe
\cite{Peacock:1996ci,Smith:2002dz,Heitmann:2008eq,Takahashi:2012em}.
Unfortunately, there are no calibrated fitting formulae of $P_{\delta}(k)$ 
for the mixed dark matter models we consider here.
We thus use direct numerical simulations
to obtain the convergence power spectra. 

\subsection{Cosmological simulations}
\subsubsection{$N$-body simulations}

It is necessary to use ray-tracing simulations in order to
study the effect of light gravitino on the weak lensing power spectrum
in nonlinear regimes.
We first run cosmological $N$-body simulations for models with light gravitinos.
We use the parallel Tree-Particle Mesh code {\tt Gadget2}
\cite{Springel:2005mi}.
Each simulation is run with $512^3$ dark matter particles in a volume of comoving 
240 ${\rm Mpc}/h$ on a side.
We generate the initial conditions following the standard Zel'dovich approximation.
We use the accurate linear matter power spectrum 
calculated by the modified {\tt CAMB} (Section\,\ref{sec:linearpower}).
It is important to generate the initial conditions
at a sufficiently low redshift so that the total matter, including the contribution
from the light gravitino, can be treated as effectively a cold component.
We set the initial redshift $z_{\rm init}=9$ because
the typical thermal velocity of the gravitino is then sufficiently small
compared to the virial velocity of the smallest halos resolved
in our simulation.
We also run a $N$-body simulation from $z_{\rm init}=49$ for the mixed 
dark matter model to
examine the overall effect caused by the choice of $z_{\rm init}$.

For our fiducial cosmology, we adopt the following parameters:
matter density $\Omega_{m}=0.3175$, dark energy density $\Omega_{\Lambda}=0.6825$
with the equation of state parameter $w_{0} = -1$,
Hubble parameter $h=0.6711$ and 
the primordial spectrum with the scalar spectral index $n_s=0.9624$
and the normalized amplitude $A_{s} = 2.215 \times 10^{-9}$
at the pivot scale $k=0.05\ {\rm Mpc}^{-1}$.
These parameters are consistent with 
the Planck mission first year results \cite{Ade:2013zuv}.
Two cases with the gravitino mass $m_{3/2}=4$ and $16$ eV
are chosen as representative models.
We summarize the simulation parameters in Table \ref{tab:nbody}.

\begin{table}
\begin{center}
\begin{tabular}{|c|c|c|c|}
\hline
& $m_{\rm 3/2}$ [eV] & $z_{\rm init}$ & \# of $N$-body sims \\ \hline
CDM     & 0 & 49 & 5 \\ \hline
MDM4--lowz  & 4 & 9 & 5 \\ \hline
MDM16--lowz & 16 & 9 & 5 \\ \hline
MDM4--highz & 4 & 49 & 5 \\ \hline
MDM16--highz & 16 & 49 & 5 \\ \hline
\end{tabular} 
\caption{Parameters for our $N$-body simulations.
	For each model, we run 5 $N$-body 
        realizations and generate 20 weak lensing convergence maps.}
\label{tab:nbody}
\end{center}
\end{table}

\subsubsection{Ray-Tracing simulation}

We generate light-cone outputs from our $N$-body simulations 
for ray-tracing simulations of gravitational lensing.
The simulation boxes are 
placed to cover a past light-cone of a hypothetical observer 
with angular extent $5^{\circ}\times 5^{\circ}$, from redshift 
$z=0$ to $z\sim 1$, similarly to the methods in  
\cite{White:1999xa, Hamana:2001vz}.
We use the standard multiple lens plane algorithm 
in order to simulate gravitational lensing signals
\cite{Jain:1999ir}.
The configuration of our simulations is similar to 
that in \cite{Sato:2009ct}.

We set the initial ray directions on $4096^2$ grids.
The corresponding angular grid size is 
$5^{\circ}/4096\sim 0.075$ arcmin.
To avoid multiple appearance of the same structure aligned 
along a line-of-sight, 
we shift randomly the $N$-body simulation boxes.
In addition, we use simulation outputs from independent
realizations when generating the light-cone outputs.
Finally we obtain 20 independent convergence maps 
from 5 $N$-body simulations for each cosmological model.
We fix the redshift of the source 
galaxies to $z_{\rm source} = 1.0$.

We measure the binned power spectrum of convergence field 
by averaging the product of Fourier modes $|\tilde{\kappa}(\bd{\ell})|^2$
for each multiple bin with $\Delta \log_{10}\ell = 0.1$ from
$\ell = 100$ to $10^{5}$.

\section{Results}
\label{sec:results}
\subsection{Convergence power spectrum}

\begin{figure}[t!]
\centering 
\includegraphics[width=.6\textwidth,clip]{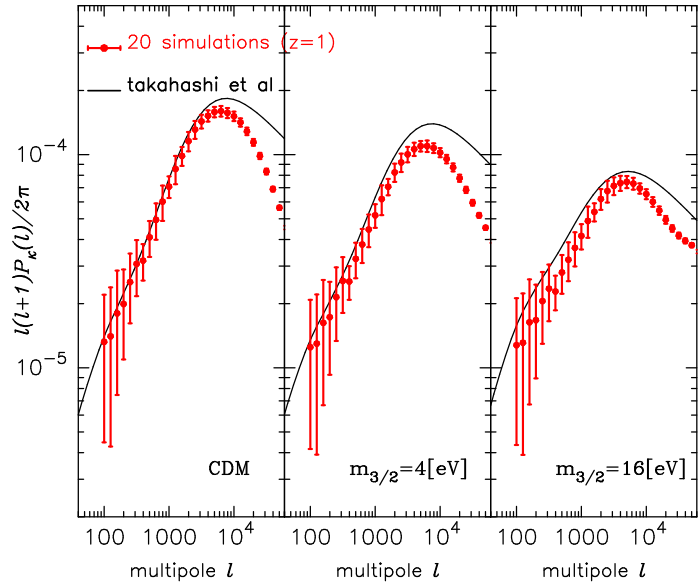}
\caption{\label{fig:llcl_comp_cdm_grav}
The convergence power spectra from our ray-tracing simulations 
for models with $m_{3/2}=0, 4$ and 16 eV are shown in the left.
In each panel, the red points represent the average measured power spectrum 
and the error bars show the standard deviation over 20 realizations.
We use the simulations that start from $z_{\rm init} = 9$ for this figure.
The solid line is calculated by eq.~\eqref{eq:kappa_power} 
and fitting formula of $P_{\delta}(k)$ in \cite{Takahashi:2012em}
with $z_{\rm source}=1.0$.}
\end{figure}

Let us first discuss how the light gravitino affects the lensing power spectrum.
Figure \ref{fig:llcl_comp_cdm_grav} compares the measured 
convergence power spectra with the analytic model prediction 
(eq.~\eqref{eq:kappa_power})
calculated by the fitting model in \cite{Takahashi:2012em}. 
The results for $m_{3/2}=0, 4$ and 16 eV are shown in the left, 
medium and right panel,
respectively. The red points show the average power spectrum 
over 20 realizations
with the error bars indicating the standard deviation of the realizations.
The solid line is the model prediction of eq.~\eqref{eq:kappa_power} 
for $z_{\rm source}=1$.
Note that the fitting function for $P_{\delta}(k)$
is calibrated for the standard $\Lambda$CDM cosmologies
with a wide range of cosmological parameters.
We thus assume 
that the non-linear evolution of $P_{\delta}(k)$ for our
mixed dark matter model is also described in the
same manner as in the
standard $\Lambda$CDM. 
In practice, we simply input the linear power spectrum for the 
mixed dark matter model (Section 3), but do not change the coefficients 
in the formula. 

We see in Figure \ref{fig:llcl_comp_cdm_grav} that 
the analytic model and the simulation result agree well to $\ell \le 4000$.
This is consistent with the results of previous 
studies \cite{Sato:2009ct,Takahashi:2012em};
the fitting model becomes less accurate  
at (sub-)arcminute scales even in the case of 
standard $\Lambda$CDM cosmology.
The convergence power spectra for the mixed dark matter model 
differ significantly from that for the $\Lambda$CDM model
even at around $\ell=1000$ corresponding to physical mega-parsec scale. 
Clearly, free-streaming of the gravitino affects the matter power spectrum
at the nonlinear scales, and thus the above simple analytic approach
does not work well even at $\ell=1000$ for the mixed dark matter model. 

\begin{figure}[t!]
\centering 
\includegraphics[width=.4\textwidth,clip]{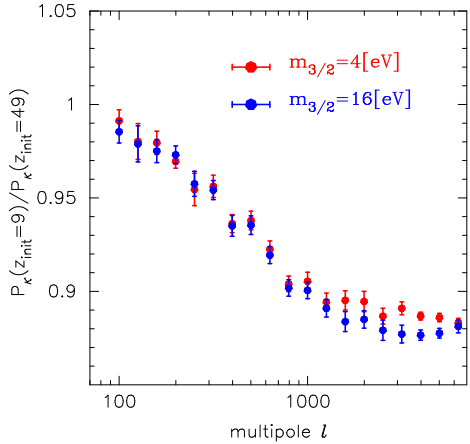}
\caption{\label{fig:model_uncertainty}
We plot the ratio of the lensing power spectra of the ray-tracing simulations 
with $z_{\rm init}=9$ and 49. The error bars indicating 
standard deviation estimated from 20 realizations.
}
\end{figure}

We have examined the effect of choice of the initial redshift.
In general, $N$-body simulations for the kind of mixed dark matter
model should be initialized at a sufficiently low redshift
in order to avoid numerical effects owing to gravitino thermal motions.
Because assigning thermal velocities to $N$-body simulation particles
is a non-trivial issue (see, e.g., \cite{AvilaReese:2000hg}), 
we do not attempt to add random velocities to the particles.
Instead, we examine how the choice of initial redshift affects
the result at low redshifts
by comparing two simulations that are started from $z_{\rm init}=9$ and 49.
Figure \ref{fig:model_uncertainty} compares the lensing
power spectra obtained from our simulations with different
$z_{\rm init}$.

The red points are for the gravitino with $m_{3/2}=4$ eV and the 
blue points for $m_{3/2}=16$ eV. Note that, unlike in ordinary
warm dark matter models, the free-streaming scale, the gravitino
mass, and the cosmic abundance are all related to each other 
in our light gravitino model.
We plot the standard deviation of mean value over 20 maps as error bars 
for each model.
We find that the initial redshift affects the convergence power spectra 
at a level of $\sim10 \ \%$.
It is important to note that the simulation from $z_{\rm init}=49$ 
is not set up consistently, 
because our simulation particles can represent 
only non-relativistic components, 
while the light gravitino is 
relativistic at such a high redshift.
Overall, Figure \ref{fig:model_uncertainty} indicates that
the simulated lensing power spectrum for the mixed dark matter model 
likely has inaccuracies with a level of $\sim 10 \%$.

\subsection{Fisher analysis}

We perform a Fisher analysis to forecast the cosmological parameter
constraints, including $m_{3/2}$.
For a multivariate Gaussian likelihood, the Fisher matrix $F_{ij}$ 
is written as
\beqa
F_{ij} = \frac{1}{2} {\rm Tr} 
\left[ A_{i} A_{j} + C^{-1} M_{ij} \right], \label{eq:Fij}
\eeqa
where $A_{i} = C^{-1} \partial C/\partial p_{i}$, 
$M_{ij} = 2 \left(\partial P_{\kappa}/\partial p_{i} \right)\left(\partial P_{\kappa}/\partial p_{j} \right)$, 
$C$ is the data covariance matrix
and $\bd{p}$ is a set of parameters of interest.
\footnote{We only consider the second term in eq.~\eqref{eq:Fij}.
Because $C$ scales approximately inverse-proportionally to survey area, 
the second term is expected to be dominant 
for a very wide area survey \cite{Eifler:2008gx}.}
In the present study, 
we choose $\bd{p}=(m_{3/2}, 10^{9}A_{s}, n_{s}, \Omega_{c}h^2, w_{0})$
as cosmological parameters to constrain.
For the Fisher analysis, we need to calculate the derivative of $P_{\kappa}$ 
with respect to $\bd{p}$.
For $m_{3/2}$, we first fit the measured power spectrum 
$P_{\kappa}(\ell)$ using a quadratic form of $m_{3/2}$, 
i.e.~$a_{0}(\ell)+a_{1}(\ell)m_{3/2}+a_{2}(\ell)m_{3/2}^2$.
We then calculate the derivative by $a_{1}(\ell)+2a_{2}(\ell)m_{3/2}$. 
For the other parameters, we evaluate the derivatives as follows:
\beqa
\frac{\partial P_{\kappa}(\ell)}{\partial p_{i}} 
= \frac{P_{\kappa}(\ell, p^{(0)}_{i}+dp_{i})-P_{\kappa}(\ell, p^{(0)}_{i}-dp_{i})}{2dp_{i}},
\label{eq:dev}
\eeqa
where $p^{(0)}_{i}$ is the fiducial value 
and $dp_{i}$ is the variation of $i$-th parameter.
Here, we simply calculate $P_{\kappa}(\ell,\bd{p})$ 
using eq.~\eqref{eq:kappa_power}
and the fitting formula of $P_{\delta}(k)$ in \cite{Takahashi:2012em}.
We summarize the fiducial values of $\bd{p}$ and $d\bd{p}$ in Table \ref{tab:param}.

\begin{table}
\begin{center}
\begin{tabular}{|c|c|c|c|c|c|}
\hline
& $m_{\rm 3/2}$ [eV] & $10^{9}A_{s}$ & $n_{s}$ & $\Omega_{c}h^2$ & $w_{0}$ \\ \hline
fiducial & 4 & 2.215 & 0.924 & 0.12029 & -1 \\ \hline
$d\bd{p}$ & -- & 0.1 & 0.01 & 0.03 & 0.1 \\ \hline
\end{tabular} 
\caption{
Parameters in our Fisher analysis.
For each parameter, we calculate the power spectra $P_{\kappa}$
with $d\bd{p}$ varied around the fiducial value
in order to calculate the derivative of Equation (\ref{eq:dev}).  
}
\label{tab:param}
\end{center}
\end{table}

The covariance matrix of the convergence power spectrum  
can be expressed as a sum of the Gaussian and non-Gaussian contributions
\cite{Cooray:2000ry, Sato:2009ct}.
Previous studies show that the non-Gaussian error degrades the constraints on
cosmological parameters with a level of $O(10 \%)$
\cite{Sato:2013mq}.
We calculate the non-Gaussian contribution  
by using 1000 lensing maps in \cite{Sato:2009ct}
in the following direct manner:
\beqa
{\rm Cov}[P_{\kappa}(\ell), P_{\kappa}(\ell^{\prime})]
=\frac{1}{N_{R}-1}\sum_{r=1}^{N_{R}}(\hat{P}^{r}_{\kappa}(\ell)-\bar{P}_{\kappa}(\ell))
(\hat{P}^{r}_{\kappa}(\ell^{\prime})-\bar{P}_{\kappa}(\ell^{\prime})), \label{eq:cov_cl}
\eeqa
where $\hat{P}_{\kappa}^{r}(\ell)$ is the measured power 
spectrum in $r$-th realization and
$\bar{P}_{\kappa}(\ell)$ is the average power spectrum 
over $N_{R} = 1000$ realizations.
The configuration of the simulation in \cite{Sato:2009ct} 
is similar to ours, which covers 25 ${\rm deg}^2$ on the sky.
When necessary, we simply scale the covariance 
matrix eq.~\eqref{eq:cov_cl} by the designated survey area.

\begin{figure}[t!]
\centering 
\includegraphics[width=.5\textwidth,clip]{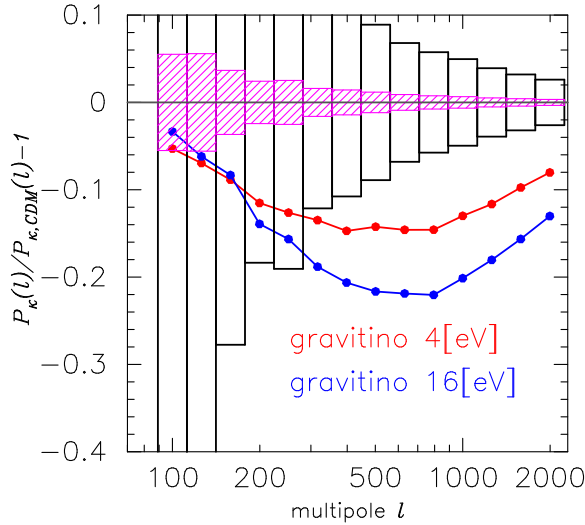}
\caption{\label{fig:pkappa_error_comp}
The derived statistical error of the lensing power spectrum.
The boxes show the statistical error of lensing power spectrum 
given by a sum of eq.~\eqref{eq:cov_cl} and ~\eqref{eq:cov_noise}.
The black boxes are for a 25 ${\rm deg}^2$ area survey,
which is same as the size of our simulation.
The purple hatched regions show the expected error 
for upcoming lensing survey with 
an area of 1500 ${\rm deg}^2$.
For comparison, we also plot the difference of the 
lensing power spectra
between the pure CDM model and mixed dark matter models.
The red line is for $m_{3/2}=4$ eV and 
the blue one for $m_{3/2}=16$ eV.
For this plot, the number density of sources 
is set to be 10 ${\rm arcmin}^{-2}$.
}
\end{figure}

We also take various systematic effects into account 
in the following manner.
It is well-known that the intrinsic 
ellipticities of source galaxies induce noises to lensing 
power spectrum. Assuming intrinsic ellipticities are 
uncorrelated, we compute the noise contribution 
to the covariance matrix of convergence power spectrum as \cite{Takada:2003hy}
\beqa
{\rm Cov}[P_{\kappa}(\ell),P_{\kappa}(\ell^{\prime})]|_{\rm noise} 
&=& \frac{2}{f_{\rm sky}(2\ell + 1)\Delta \ell}P_{\rm noise}
\left(P_{\rm noise}+2P_{\kappa}(\ell)\right) \delta_{\ell \ell^{\prime}},\label{eq:cov_noise} \\
P_{\rm noise}
&=& \frac{1}{n_{\rm gal}}\left(\frac{\sigma_{\rm int}}{\cal R}\right)^2, 
\label{eq:noise}
\eeqa
where 
$\Delta \ell$ is the width of the multipole bin,
$f_{\rm sky}$ is the fraction of sky covered,
$n_{\rm gal}$ is the number density of source galaxies,
${\cal R}$ is the shear response,
and $\sigma_{\rm int}$ is the root-mean-square of the shear noise.
Throughout the present paper, we adopt ${\cal R}=1.7$ and $\sigma_{\rm int}=0.4$.
The values are typical in ground based weak lensing surveys
\cite{Miyazaki:2002xu, Mandelbaum:2005nx}.
We finally obtain the covariance matrix for our Fisher analysis 
as a sum of eq.~\eqref{eq:cov_cl} and \eqref{eq:cov_noise}.
In Figure \ref{fig:pkappa_error_comp}, we compare the derived statistical 
error (the square root of the diagonal part of the covariance matrix)
and the estimated difference of the lensing power spectra between the 
mixed dark matter models considered here.
Clearly, future wide field lensing surveys with 1500 
square degrees can discriminate (or constrain)
the light gravitino models. There are some certain degeneracies
among the cosmological parameters, which we shall discuss
in section 5.4.

We explore more realistic constraints by using priors 
expected from the cosmological parameter estimates 
from the Planck satellite mission.
When we compute the Fisher matrix for the CMB, we use 
the Markov-Chain Monte-Carlo (MCMC) engine 
{\tt COSMOMC} \cite{Lewis:2002ah} for exploring cosmological 
parameter space.
We consider the parameter constraints from the angular 
power spectra of temperature anisotropies, 
$E$-mode polarization, and 
their cross-correlation.
For MCMC, in addition to $10^{9}A_{s}, n_{s}, \Omega_{c}h^2$ and $w_{0}$, 
we adopt the baryon density $\Omega_{b} h^2$, 
Hubble parameter $h$, and
reionization optical depth $\tau$ as independent parameters.
To examine the potential of lensing power spectrum  
to constrain $m_{3/2}$, 
we do not assume any prior on $m_{3/2}$ from the CMB.
Assuming that the constraints from the CMB and the 
lensing power spectrum are independent of each other, 
we express the total Fisher matrix as
\beqa
\bd{F} = \bd{F}_{\rm lensing} + \bd{F}_{\rm CMB}. \label{eq:Ftot}
\eeqa
When we include the CMB priors in this way,
we marginalize over the other cosmological parameters 
except $\bd{p}=(m_{3/2}, 10^{9}A_{s}, n_{s}, \Omega_{c}h^2, w_{0})$.

\subsection{Forecast for future surveys}

\begin{figure}[t!]
\centering 
\includegraphics[width=.5\textwidth,clip]{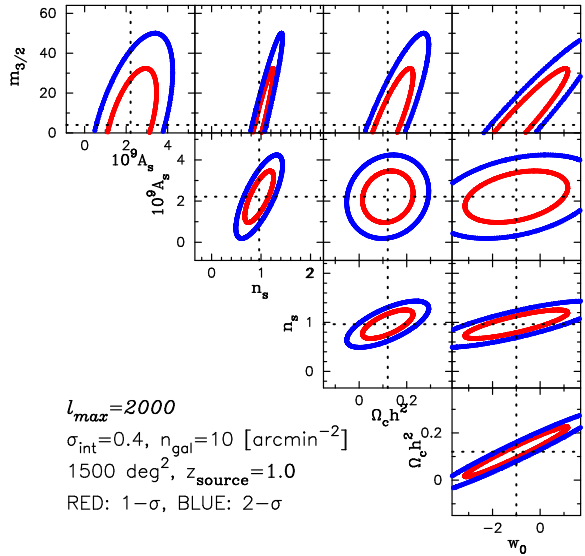}
\caption{\label{fig:error_cir}
We show the cosmological 
constraints from lensing power spectrum alone.
We consider the upcoming Subaru Hyper Suprime-Cam survey 
with an area of 1500 ${\rm deg}^2$. 
}
\end{figure}

We provide the forecast for upcoming weak lensing surveys 
with an area coverage of more than
a thousand square degrees.
We use logarithmically spaced bins with $\Delta \log_{10}\ell = 0.1$ from
$\ell = 100$ to 2000.
We thus need a $14 \times 14$ covariance matrix of 
lensing power spectrum in the Fisher analysis.
Our 1000 mock observations are sufficiently 
large to estimate the covariance matrix accurately.

\begin{figure}[t!]
\centering 
\includegraphics[width=.5\textwidth,clip]{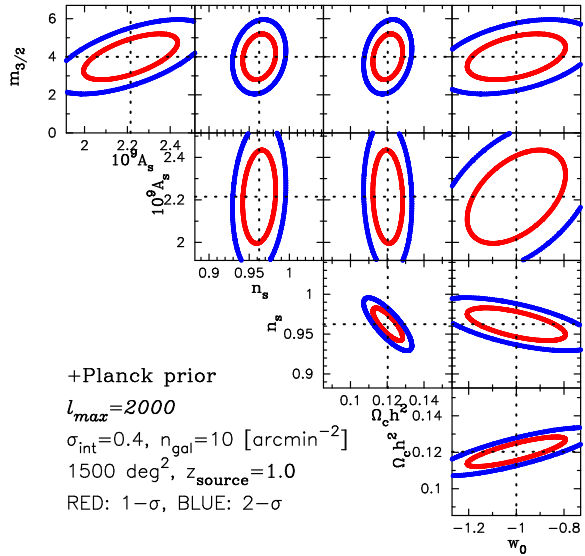}
\caption{\label{fig:error_cir+cmb}
As for Figure \ref{fig:error_cir}, but with the CMB priors
described in Section 5.2.
}
\end{figure}

Figure \ref{fig:error_cir} shows the two-dimensional 
confidence contours for the Subaru Hyper Suprime-Cam (HSC)
lensing survey 
\footnote{\rm{http://www.naoj.org/Projects/HSC/j\_index.html}}.
We assume $n_{\rm gal}=10\ {\rm arcmin}^{-2}$.
The red circles show the constraints with 68 \% confidence 
level (1$\sigma$) whereas the blue ones correspond to
95 \% confidence level (2$\sigma$).
The marginalized 1$\sigma$ error 
for $m_{3/2}$ over other parameters is found to be $\sim 18$ eV.
Note that this is a constraint from the lensing survey alone.
We also show the forecast with the CMB priors 
in Figure \ref{fig:error_cir+cmb}.
The constraint on the gravitino mass 
is significantly improved in this case, because 
using the CMB data breaks some 
degeneracies among cosmological parameters, 
e.g. $10^9A_{s}$ and $\Omega_{c}h^2$
\cite{Kilbinger:2012qz}.
It is impressive that we can derive constraint on the gravitino 
mass with a level of 1 eV by combining data from the HSC lensing survey 
and the Planck mission.

\subsection{Degeneracy between massive neutrino}

It is well known that massive neutrinos affect the lensing power spectrum 
in a similar way to the light gravitino; 
free-streaming of massive neutrinos suppress the growth of structure.
At large length scales, the effect on $P_{\delta}(k)$ has been quantified 
by linear theory and extensions to first-order perturbation theory, 
e.g.~\cite{Bond:1980ha, Saito:2008bp}.
Probing the effect of massive neutrinos on $P_{\delta}(k)$ 
in the fully non-linear regime is still challenging,
because it is difficult to include the relativistic species
in $N$-body simulations 
\cite{Brandbyge:2008js,Brandbyge:2008rv,Viel:2010bn,Bird:2011rb}.
In order to study the degeneracy between the light gravitino mass 
and the total mass of massive neutrinos in the cosmological
parameter estimate,
we utilize a fitting model of 
$P_{\delta}(k)$ that include the effect of neutrinos \cite{Bird:2011rb}.

\begin{figure}[t!]
\centering 
\includegraphics[width=.4\textwidth,clip]{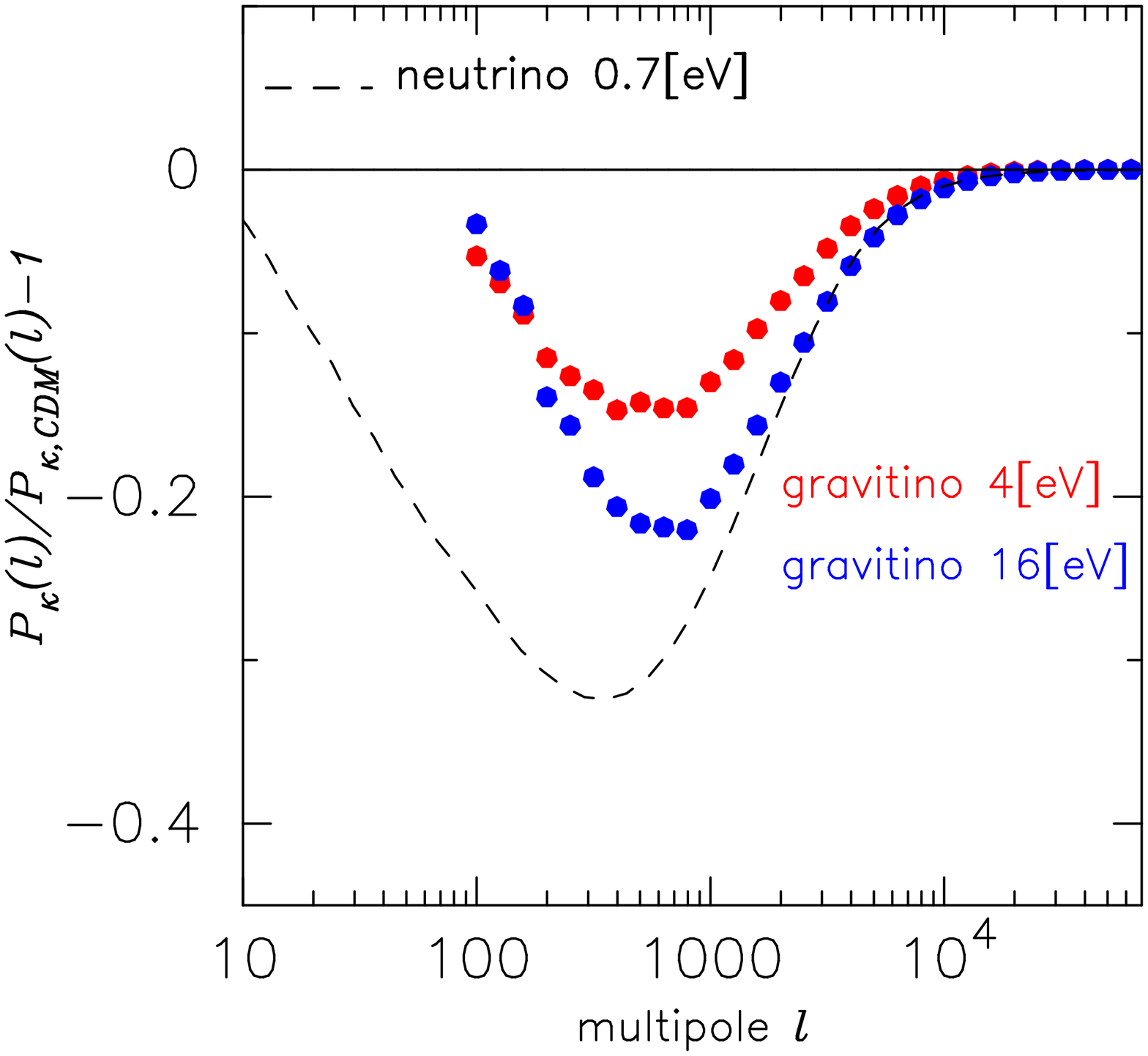}
\includegraphics[width=.4\textwidth,clip]{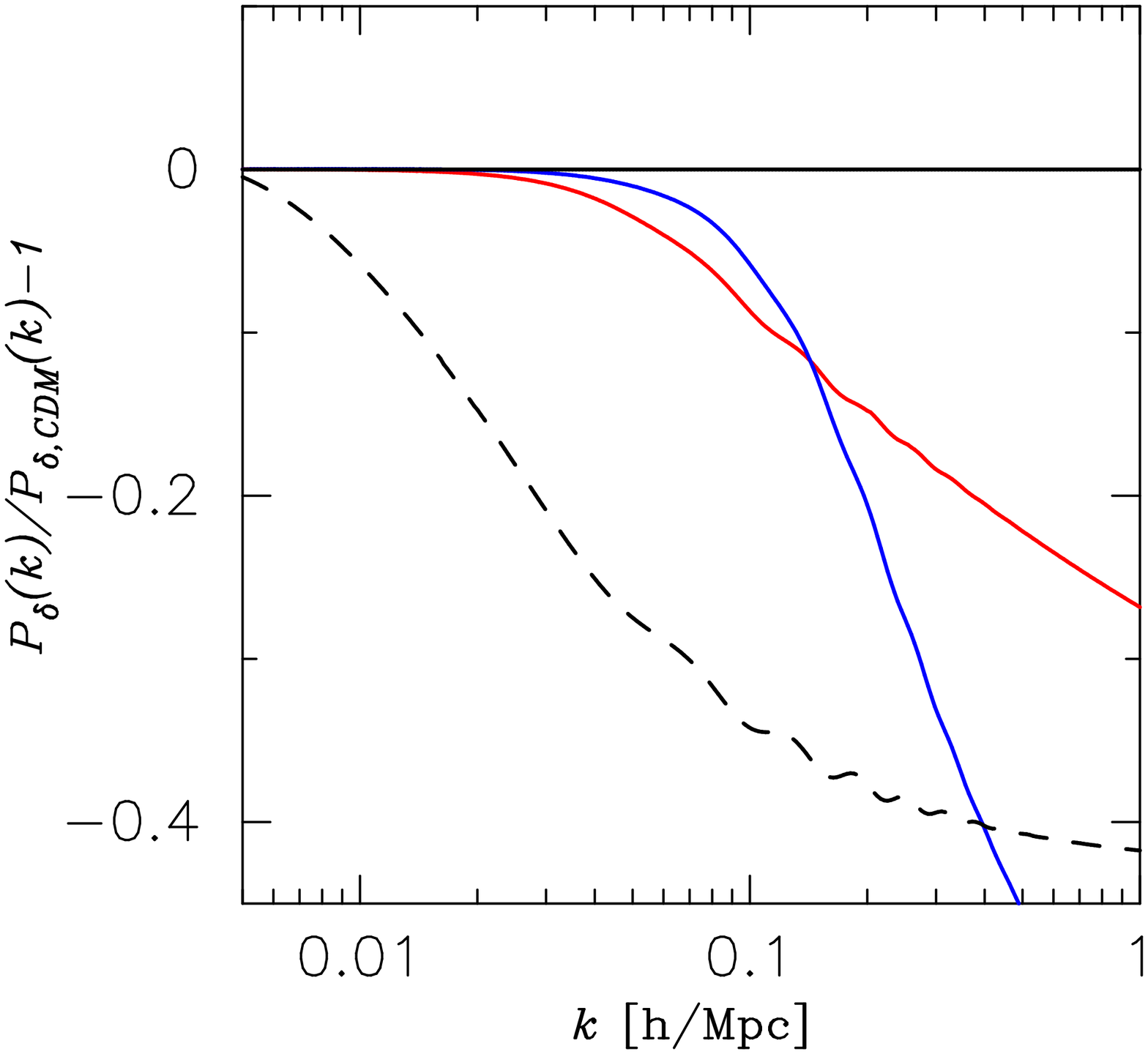}
\caption{\label{fig:llcl_comp_cdm_grav_nu}
The effect of the light gravitino and massive neutrinos on the lensing power 
spectrum (left) and on the three dimensional matter power spectrum 
at $z=0$ (right).
In each panel, the dashed line shows the resulting lensing power 
spectrum calculated by the fitting model in
\cite{Bird:2011rb} including the effect	of massive neutrinos.
We assume $m_{\nu, {\rm tot}}=0.7$ eV.
The points show the measured power spectra from our 
simulations in the case of $m_{3/2}=4$ eV (red) 
and $m_{3/2}=16$ eV (blue). 
We show these power spectra normalized by that for the pure CDM model.
}
\end{figure}

Figure \ref{fig:llcl_comp_cdm_grav_nu} shows the effect of massive 
neutrinos on the lensing power spectrum.
There, we assume the mass of neutrino $m_{\nu, {\rm tot}}=0.7$ eV, 
which is comparable to the current upper limits 
with 95 percent confidence 
\cite{Reid:2009xm, Saito:2010pw, Zhao:2012xw}.
We compare the lensing power spectrum with those 
of the light gravitino with $m_{3/2}=4$ and 16 eV.
As expected, massive neutrinos with $m_{\nu, {\rm tot}}=0.7$ eV 
causes a similar effect on the lensing power spectrum
to that of the light gravitino.
It would thus be difficult to break the degeneracy 
between the contribution of the light gravitino and 
that of massive neutrinos by a weak lensing survey alone.
We need other probes of the matter distribution at large scales 
and at different epochs, such as galaxy clustering.
For example, future galaxy redshift surveys are aimed at measuring the 
galaxy clustering at $k\sim 0.01 - 0.1$ $h/{\rm Mpc}$.
At the quasi-nonlinear scales, the effect of massive neutrinos
on $P_{\delta}(k)$ can be distinguishable
from that of the light gravitino, 
as shown in the right panel of 
Figure \ref{fig:llcl_comp_cdm_grav_nu}.

\section{Summary and Discussion}
\label{seq:conclusion}

The gravitino mass is one of the fundamental parameters in SUSY theory
that is directly related to the SUSY breaking energy scale.
We focus on the gauge-mediated supersymmetry breaking model
that generically predicts the existence of light gravitinos with 
$m_{3/2} \sim$\,eV-keV.
Free-streaming of such light gravitino affects the matter distribution
significantly, leaving characteristic 
suppression in the matter power spectrum at around 
$k \gtrsim 0.1$\,h/Mpc.
The nonlinear scale is beyond the reach of 
the CMB anisotropy measurements. 
We show that observations of weak gravitational lensing  
can be used to probe the matter distribution at the relevant
length scales and thus can be used 
to detect the imprints of the light gravitino.

We have explored cosmological constraints on the light gravitino mass 
from cosmic shear statistics.
Our ray-tracing simulations have revealed that the conventional model 
for nonlinear 
correction to the matter power spectrum 
\cite{Takahashi:2012em} does not work well for 
models with the light gravitino.
The difference between the simulation results and the fitting formula 
is significant at $\ell \sim 1000$, where upcoming 
lensing surveys are aimed at measuring the power spectrum accurately.
Using a large set of ray-tracing simulations,
we have shown that the HSC like survey has a potential to determine 
the gravitino mass 
with an accuracy of $4\pm1$\,eV
with the help of Planck CMB priors on the basic cosmological parameters.

\begin{figure}[t!]
\centering 
\includegraphics[width=.45\textwidth,clip]{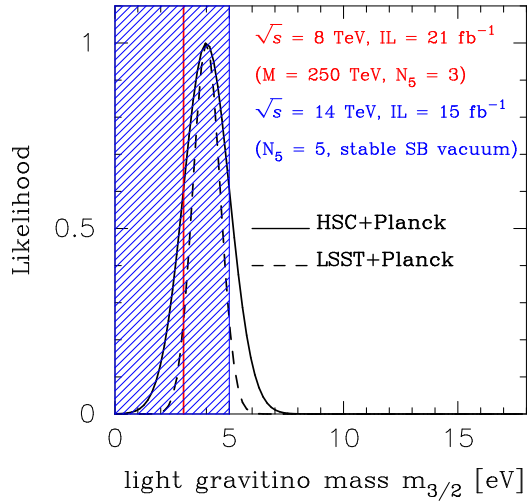}
\caption{\label{fig:L_mass}
The likelihood distribution of $m_{3/2}$ expected by future weak 
lensing surveys.
We have used the binned lensing power spectrum with the CMB prior 
for this figure.
The solid line corresponds to the Hyper Suprime-Cam survey
and the dashed one is for Large Synoptic Survey Telescope.
The vertical lines show the current/future focus points 
of the GMSB model at the LHC (Table\,\ref{table:lowerbound}).
In the near-future LHC, all GMSB models with $m_{3/2}<5$\,eV 
(shaded region) can be probed 
if they involve the stable SUSY breaking (SB) vacuum 
and the successful grand unification.
}
\end{figure}

Let us further discuss prospects for future lensing surveys.
For the upcoming survey with 20000 ${\rm deg}^2$ 
by the Large Synoptic Survey Telescope (LSST)
\footnote{\rm{http://www.lsst.org/lsst/}}, 
we will be able to use fainter galaxies for lensing analysis. 
Effectively the number of source galaxies will be larger.
In this case, we can constrain on $m_{3/2}$ with a level of 
$4\pm0.6$ eV assuming $n_{\rm gal}=15$ ${\rm arcmin}^{-2}$.
Note that the constraint is tighter than the current one 
from the Lyman-$\alpha$ 
forest by a factor of $\sim 10$
\cite{Viel:2005qj}, 
and also comparable to the forecast that utilizes CMB lensing
\cite{Ichikawa:2009ir}.
Although there is some certain degeneracy between the effect of 
massive neutrinos and that of the light 
gravitino in cosmic shear power spectrum,
we argue that combining cosmic 
shear and galaxy clustering and/or CMB lensing can break the degeneracy.
We summarize the forecast of $m_{3/2}$ from upcoming weak lensing 
survey in figure \ref{fig:L_mass}.
There, we estimate the future constraint on the gravitino mass 
from the LHC $14$\,TeV run for the minimal GMSB model.

Ultimately, the International Linear Collider (ILC) experiment 
has a potential to determine the gravitino mass.
When the next lightest supersymmetric particle is stau, 
its lifetime is proportional to the gravitino mass squared.
By measuring the distribution of the impact parameter, 
one can evaluate the stau lifetime and hence 
the gravitino mass\,\cite{Matsumoto:2011fk}. %
\footnote{
To this end, the center of mass energy should exceed two time 
the stau mass and the background events should be effectively 
eliminated.
However, this may be challenging for the present design of 
the ILC in the case of the heavy stau for a given gravitino 
mass, i.e. large $N_{5}$ and $\lambda x$.}
Overall, under reasonable assumptions, 
almost all of the interesting GMSB models with $m_{3/2}<5$\,eV can be 
probed in $15\,{\rm fb}^{-1}$ of LHC $14$\,TeV run.
Combining cosmological and collider searches together, 
we will reach the conclusion about the GMSB model. 

\acknowledgments
A.K. would like to thank M. Ibe, S. Matsumoto, and T. T. Yanagida for useful discussions.
M.S. appreciates M. Sato for kindly providing us the set of simulations.
Numerical computations for the present work have been carried 
out on Cray XC30 at Center for Computational Astrophysics, CfCA, 
of National Astronomical Observatory of Japan, and in part under 
the Interdisciplinary Computational Science Program in Center for Computational Sciences, 
University of Tsukuba.
This work is supported in part by JSPS Research Fellowship for Young Scientists (A.K. and M.S.) and by World Premier International Research Center Initiative, MEXT, Japan.

\bibliographystyle{JHEP}
\bibliography{ref}

\end{document}